\documentclass[12pt,letter]{article}
\pdfoutput=1
\usepackage{graphicx, epsfig, color,cite}
\usepackage{amsmath}
\usepackage{amssymb}
\usepackage{float}
\usepackage{subfig}
\usepackage{hyperref}

\def\eslt{\not\!\!\!{E_T}}
\def\to{\rightarrow}

\def\bi{\begin{itemize}}
\def\ei{\end{itemize}}

\def\tchi{\tilde\chi}

\def\tst{\tilde t}
\def\ttau{\tilde \tau}

\def\tnu{\tilde\nu}

\def\alt{\lesssim}
\def\agt{\gtrsim}
\def\be{\begin{equation}}  
\def\ee{\end{equation}}  
\def\bea{\begin{eqnarray}}  
\def\eea{\end{eqnarray}}

\begin{document}
\begin{titlepage}
\begin{flushright}
OU-HEP-240328
\end{flushright}

\vspace{0.5cm}
\begin{center}
  {\Large \bf Stau pairs from natural SUSY\\ at high luminosity LHC}\\
\vspace{1.2cm} \renewcommand{\thefootnote}{\fnsymbol{footnote}}
{\large Howard Baer$^{1}$\footnote[1]{Email: baer@ou.edu },
Vernon Barger$^2$\footnote[2]{Email: barger@pheno.wisc.edu} and
Kairui Zhang$^2$\footnote[5]{Email: kzhang89@wisc.edu}
}\\ 
\vspace{1.2cm} \renewcommand{\thefootnote}{\arabic{footnote}}
{\it 
$^1$Homer L. Dodge Department of Physics and Astronomy,
University of Oklahoma, Norman, OK 73019, USA \\[3pt]
}
{\it 
$^2$Department of Physics,
University of Wisconsin, Madison, WI 53706 USA \\[3pt]
}

\end{center}

\vspace{0.5cm}
\begin{abstract}
\noindent
Natural supersymmetry (SUSY) with light higgsinos is perhaps the most
plausible of all weak scale SUSY models while a variety of motivations
point to (right) tau sleptons as the lightest of all the sleptons.
We examine a SUSY model line with rather light right-staus embedded
within natural SUSY.
For light $\ttau_1$ of a few hundred GeV, then the decays
$\ttau_1\to \tau\tchi_{1,2}^0$ and $\nu_\tau\tchi_1^-$ occur
at comparable rates where the (higgsino-like)
$\tchi_1^\pm$ and $\tchi_2^0$ release only small visible energy:
in this case, the expected $\tau^+\tau^- +\eslt$ signature is diminished
from usual expectations due to the presence of the nearly invisible
decay mode $\ttau_1\to\nu_\tau\tchi_1^-$. However, once $m_{\ttau_1}\agt m(bino)$,
then decays to binos such as $\ttau_1\to \tau\tchi_3^0$ open up where
$\tchi_3^0$ decays to higgsinos plus $W^\pm$, $Z^0$ and $h$
at comparable rates. For these heavier staus, then stau pair production
gives rise to diboson$+\eslt$ events which may contain 0, 1 or 2 additional
hard $\tau$ leptons.
From these considerations, we examine the potential for future discovery
of tau-slepton pair production at high-luminosity LHC.
While we do not find a $5\sigma$ HL-LHC discovery reach for 3000 fb$^{-1}$,
we do find a 95\% CL exclusion reach, ranging between
$m_{\ttau_1}:100-450$ GeV for $m_{\tchi_1^0}\sim 100$ GeV.
This latter reach disappears for $m_{\tchi_1^0}\agt 200$ GeV.

\end{abstract}
\end{titlepage}

\section{Introduction}
\label{sec:intro}

The search for weak scale supersymmetry (SUSY)\cite{Baer:2006rs} at hadron collider
experiments often focusses on strongly interacting sparticles--
the gluinos and squarks--
since these have the largest production cross section for a given
sparticle mass value. However, in many models the gluinos and squarks
also have the largest mass values, while sleptons and electroweakinos (EWinos)
are much lighter. Thus, for a given point in model parameter space,
sleptons and EWino pair production may dominate the production cross sections
by virtue of their smaller mass values.

In a previous work\cite{Baer:2023olq}, we examined prospects for EWino
pair production at luminosity upgrades of the CERN Large Hadron Collider
(LHC) in the context of natural SUSY which is characterized by low values of {\it electroweak} finetuning
measure $\Delta_{EW}\alt 30$\cite{Baer:2012up,Baer:2012cf}.
The value of $\Delta_{EW}$ is a measure of practical naturalness\cite{Baer:2023cvi}:
that all independent contributions to an observable should be comparable to or
less than its measured value.
For the case of the Minimal Supersymmetric Standard Model, or MSSM, the weak scale
(as typified by the $Z$-boson mass) is related to weak scale SUSY Lagrangian parameters as
\be
m_Z^2/2=\frac{(m_{H_d}^2+\Sigma_d^d)-(m_{H_u}^2+\Sigma_u^u)\tan^2\beta}{\tan^2\beta -1}-\mu^2\simeq -m_{H_u}^2-\mu^2-\Sigma_u^u(\tst_{1,2})
\label{eq:mzs}
\ee
where $m_{H_{u,d}}^2$ are soft breaking squared masses of the Higgs doublets,
$\tan\beta =v_u/v_d$
is the ratio of Higgs field vevs, $\mu$ is the SUSY conserving $\mu$
parameter\footnote{Twenty solutions to the SUSY $\mu$ problem
  are reviewed in Ref. \cite{Bae:2019dgg}.}
and the $\Sigma_{u,d}^{u,d}$ terms contain over 40 1-loop and some 2-loop
corrections to the scalar potential (explicit expressions are given in
Ref's \cite{Baer:2012cf,Baer:2021tta}).
The measure $\Delta_{EW}$ is defined as
\be
\Delta_{EW}\equiv |{\rm largest\ term\ on\ RHS\ of\ Eq. \ref{eq:mzs}}|/(m_Z^2/2)
  \label{eq:dew}
  \ee
so that no large unnatural finetunings are allowed in the derivation of $m_Z$.
A large negative value of $A_t$ which enters the expressions for
$\Sigma_u^u(\tst_{1,2})$ leads to large cancellations (more naturalness)
in both these terms\cite{Baer:2012up} whilst lifting $m_h\to 125$ GeV\cite{Slavich:2020zjv}.

Natural SUSY (natSUSY) models are considered more plausible than unnatural
models in that they contain no implausible accidental tunings of unrelated terms
in Eq. \ref{eq:mzs}.
Furthermore, it is now understood that natSUSY is the most likely expression
of weak scale SUSY that ought to emerge from the string
landscape\cite{Baer:2022wxe,Baer:2022dfc,Baer:2024kms}.
This arises since low $\Delta_{EW}\alt 30$
corresponds well with the anthropic Agrawal-Barr-Donoghue-Seckel (ABDS)\cite{Agrawal:1997gf,Agrawal:1998xa}
window of allowed weak scale values which give rise to
complexity in the multiverse (atomic principle).
For the case of fine-tuned models, then the available
multiverse scan space shrinks to tiny volumes compared to natural models
due to the finetuning which is required.

In the present work, we examine prospects for slepton
(specifically, the lightest $\tau$-slepton $\ttau_1$) pair production
at the high-luminosity upgrade of LHC (HL-LHC) in a natural SUSY context.
We focus on lightest right-tau-sleptons for several reasons.

In models with high scale slepton universality
(such as mSUGRA\cite{Barger:1992ac,Barger:1993gh,Arnowitt:1993qp}/CMSSM\cite{Kane:1993td}
or NUHM2, NUHM3 or NUHM4\cite{Ellis:2002wv,Baer:2005bu} models),
the stau soft mass RGEs are given by
\bea
\frac{dm_{L_3}^2}{dt}& =&\frac{2}{16\pi^2}\left( -\frac{3}{5}g_1^2M_1^2
-3g_2^2M_2^2-\frac{3}{10}g_1^2 S+f_\tau^2 X_\tau\right), \label{eq:mL3}\\
\frac{dm_{E_3}^2}{dt}& =&\frac{2}{16\pi^2}\left(-\frac{12}{5}g_1^2M_1^2
+\frac{3}{5}g_1^2 S+2f_\tau^2X_\tau\right) ,\label{eq:mE3}
\eea
where $m_{L_3}^2$ is the third generation doublet slepton soft mass
squared (giving rise to left-staus) and $m_{E_3}^2$ is the corresponding
$SU(2)_L$ singlet slepton mass squared (giving rise to right-staus).
Also, $S=m_{H_u}^2-m_{H_d}^2+Tr\left[{\bf m}_Q^2-{\bf m}_L^2-2{\bf m}_U^2+{\bf m}_D^2+{\bf m}_E^2\right]$ and $X_\tau =m_{L_3}^2+m_{E_3}^2+m_{H_d}^2+A_\tau^2$
and $t=\log (Q)$. When running from high scales ({\it e.g.} $Q=m_{GUT}$)
to $Q=m_{weak}$, the $SU(2)_L$ gauge term in Eq.~\ref{eq:mL3} drives
$m_{L_3}^2$ to larger values than $m_{E_3}^2$ at the weak scale while the
rather large $\tau$-Yukawa coupling term containing $2 f_\tau^2$
in Eq.~\ref{eq:mE3} drives the right-stau soft mass squared $m_{E_3}^2$
to smaller values than $m_{L_3}^2$.
For the natSUSY models considered here, usually $S>0$ so this term also drives
right-sleptons to smaller masses than left-sleptons at the weak scale.
Thus, in models with intra-generation
universality of scalar masses (which are motivated by $SO(10)$ where all
elements of each generation live in a single ${\bf 16}-d$ spinor rep),
we expect that right-stau masses are smaller than left-stau masses.\footnote{
  Light tau-sleptons also arise in supersymmetric twin-Higgs models: see
  {\it e.g.} Ref's \cite{Badziak:2017wxn,Badziak:2019zys,Badziak:2023gpz}.}

Also, on the theory side, the string landscape pulls soft breaking terms
as large as possible until they over-contribute beyond the ABDS window
to the weak scale. This effect tends to pull first/second generation
sfermion masses to the tens-of-TeV values whilst third generation
sfermions, which contribute proportional to their Yukawa couplings squared,
only get pulled up to values of several TeV\cite{Baer:2017uvn}
at the high scale.

Furthermore, in orbifold compactifications on the
minilandscape\cite{Lebedev:2006kn}, first/second
generation sfermions live near orbifold fixed points and
``feel much less supersymmetry than third generation fields''\cite{Nilles:2014owa} which instead live more in the
bulk where they have large overlap with Higgs multiplets.
Thus, third generation soft terms are more protected by SUSY and hence
gain smaller soft masses than their first/second generation counterparts.

On the phenomenology side, light sleptons are preferred by the $(g-2)_\mu$
anomaly\cite{Feng:2001tr}, and of all the sleptons,
the right-staus are expected to be lightest.
Also, light tau-sleptons with mass $m_{\ttau_1}\sim m_{\tchi_1^0}$
are required to {\it thermally} match the measured dark matter relic density
in the so-called stau coannihilation region of SUSY model
parameter space\cite{Ellis:1998kh,Ellis:1999mm,Baer:2002fv}.

For these reasons, we examine prospects for detecting the lightest (right-)
tau-sleptons, but within the context of natSUSY models.
To this end, in Sec. \ref{sec:model}, we develop a natural SUSY model
line with low $\Delta_{EW}$ but with a variable right-stau soft mass.
In Sec. \ref{sec:prod}, we present stau pair production cross sections
which are expected at LHC14, and in Sec. \ref{sec:decay}, we compute expected
$\ttau_1$ branching fractions along our model line. Since we work within a
natural SUSY context, then higgsinos are expected with mass
$\sim \mu\sim 100-350$ GeV\cite{Chan:1997bi,Baer:2011ec}.
The presence of light higgsinos is expected to diminish the LHC reach
for light staus compared to usual simplified models
in that in the natSUSY case, a substantial branching fraction
$\ttau_1\to\nu_\tau\tchi_1^-$ where the $\tchi_1^\pm \to f\bar{f}'\tchi_1^0$
and the small $m_{\tchi_1^+}-m_{\tchi_1^0}$ mass gap leads to very low energy
visible decay products. However, in the case where
$m_{\ttau_1}>m(bino)\gg m(higgsino)$, then
the decay to binos rapidly dominates the stau decay rate leading to
possibly new discovery signatures: diboson$+\tau\bar{\tau}+\eslt$.
In Sec. \ref{sec:reach}, we evaluate the projected HL-LHC reach
for stau pairs in natSUSY in the $m_{\ttau_1}$ vs. $m_{\tchi_1^0}$ plane.
While we do not find a $5\sigma$ discovery reach for HL-LHC, we do find
a 95\%CL exclusion reach that extends from $m_{\ttau_1}\sim 200-450$ GeV
for $m_{\tchi_1^0}\sim 100$ GeV.
We conclude in Sec. \ref{sec:conclude}.

\subsection{Brief review of some previous works}

Many early works were focussed on stau pair production in the stau
coannihilation region of models like mSUGRA/CMSSM\cite{Arnowitt:1993qp,Kane:1993td} with $\mu\gg m_Z$
and with a bino-like LSP\cite{Arnowitt:2006jq,Arnowitt:2007nt,Desai:2014uha,Berggren:2015qua,Florez:2016lwi,Aboubrahim:2017aen,Chakraborti:2023pis}.\footnote{
  Production of lighter stau pairs from heavy Higgs decay has been considered
  in Ref's \cite{Arganda:2018hdn} and \cite{Arganda:2021qgi}.}
Such models are nowadays regarded as
unnatural under $\Delta_{EW}$ and hence rather
implausible\cite{Baer:2012mv,Baer:2013gva,Baer:2014ica} as a realization
of weak scale SUSY.

In contrast, natural SUSY models with $\mu\sim m_Z$ contain three light
higgsinos $\tchi_{1,2}^0$ and $\tchi_1^\pm$.
Since higgsinos annihilate and coannihilate at high rates in the early
universe\cite{Baer:2010wm} they have no
dark matter overproduction problem and hence no need for tuning
the relic abundance into the stau coannihilation region and
there is no reason to expect a situation with
long-lived light staus. Instead, any light staus are expected to decay
promptly to the three light higgsinos along with tau-leptons or
tau-neutrinos (see upcoming Fig. \ref{fig:BFs} for branching fractions).

In Ref. \cite{ATLAS:2019gti}, the ATLAS collaboration reported on a search
for stau pair production using Run 2 data with 139 fb$^{-1}$ at
$\sqrt{s}=13$ TeV. The final state signature searched for was
$\tau_h\tau_h+\eslt$ where $\tau_h$ is a hadronically-decaying $\tau$-jet.
No signal was seen above SM backgrounds, and limits were placed in the
$m_{\ttau}$ vs. $m_{\tchi_1^0}$ plane assuming
\begin{enumerate}
\item degenerate left- and right tau-sleptons and
\item  just pair production of left tau-sleptons.
\end{enumerate}
In both cases, a simplified model with the decay $\ttau_1\to\tau\tchi_1^0$
at 100\% branching fraction was assumed.
In the former Case, with $m_{\tchi_1^0}\sim 100$ GeV, then
$m_{\ttau_{L/R}}: 230\ {\rm GeV}-350\ {\rm GeV}$ was excluded while for the
second Case no limit ensues for $m_{\tchi_1^0}=100$ GeV although
$m_{\ttau_{L}}: 150\ {\rm GeV}-320\ {\rm GeV}$ can be excluded for
$m_{\tchi_1^0}=0$.

A similar search was reported on by CMS using Run 2 data with 138 fb$^{-1}$
in Ref. \cite{CMS:2022syk}.
No excess was seen in the $\tau_h\tau_h+\eslt$ signal channel above SM
background leading CMS in Case 1 to exclude  
$m_{\ttau_{L/R}}: 200\ {\rm GeV}-380\ {\rm GeV}$ while in Case 2
$m_{\ttau_L}\sim 280$ GeV could be exluded for $m_{\tchi_1^0}=100$ GeV
while $m_{\ttau_{L}}: 120\ {\rm GeV}-350\ {\rm GeV}$ could be excluded for
$m_{\tchi_1^0}=0$.

The ATLAS Collaboration also performed a HL-LHC reach study in
2016\cite{ATLAS:2016fhh} for stau pair production with $\sqrt{s}=14$ TeV
assuming 3000 fb$^{-1}$ of integrated luminosity.
This study examined the reach for Case 1 and 2
as above but also included reach results for just
$\ttau_R\bar{\ttau}_R$ pair production (Case 3). For $m_{\tchi_1^0}=100$
GeV, they report a 95\% CL exclusion reach in $m_{\ttau}$ up to
540 GeV in Case 3, 650 GeV in case 2 and 700 GeV in case 1.
There was no $5\sigma$ discovery reach for any value of $m_{\ttau}$ in Case 3.
An updated 2018 study by ATLAS was presented in Ref. \cite{ATLAS:2018diz}.
A similar study was performed by CMS in 2019\cite{CMS:2018imu}
where a 95\% CL exclusion reach for HL-LHC in $m_{\ttau}$ up to 640 GeV
was reported for Case 1 with $m_{\tchi_1^0}=100$ GeV.

\section{A natural SUSY model line with light right-staus}
\label{sec:model}

We would like to embed light tau-sleptons within a natural SUSY model
framework since it can be argued that SUSY models with low
electroweak finetuning (with $\Delta_{EW}\alt 30$) are the most plausible
of SUSY models in that the weak scale $m_{weak}\simeq m_{W,Z,h}$ 
is of order $\sim 100$ GeV because all MSSM contributions
(some positive, some negative) in Eq. \ref{eq:mzs} are comparable
(within a factor of several) to the $m_{weak}$ scale.
Such models can be found for instance within the framework of
non-universal Higgs models\cite{Ellis:2002wv,Ellis:2002iu,Baer:2005bu}.
Here, we will work within the NUHM4 model\footnote{Four extra parameter
  non-universal Higgs model, where the four extra parameters beyond CMSSM
  include $m_0(2)$, $m_0(3)$, $m_{H_u}$ and $m_{H_d}$.}
with parameters
\be
m_0(i),\ m_{1/2},\ A_0,\ \tan\beta,\ m_{H_u},\ m_{H_d}\ \ \ (NUHM4^\prime)
\ee
(where $m_0(i)$ refers to separate soft masses $m_0(1,2,3)$ for each generation,
as is expected from general supergravity models where no known symmetry
enforces generational mass universality\cite{Soni:1983rm,Kaplunovsky:1993rd,Brignole:1993dj})
and where it is common to trade the high scale soft terms $m_{H_u}^2$
and $m_{H_d}^2$ for the more convenient weak scale parameters
$\mu$ and $m_A$:
\be
m_0(i),\ m_{1/2},\ A_0,\ \tan\beta,\ \mu,\ m_{A}\ \ \ (NUHM4)
\ee
and where $i=1-3$ is a generation index. In NUHM models, the required
$\mu\sim 100-350$ GeV parameter can be dialed to fulfill one of
the requirements of low $\Delta_{EW}$ in Eq. \ref{eq:mzs}.
Also, a large negative $A_0$
parameter lifts $m_h\to 125$ GeV\cite{Baer:2011ab,Slavich:2020zjv} while reducing
the top-squark loop corrections $\Sigma_u^u (\tst_{1,2})$ to
Eq. \ref{eq:mzs}\cite{Baer:2012up,Baer:2012cf}.
This latter effect reconciles natural SUSY with the rather
large measured value of $m_h$ and with $m_{\tst}\sim 1-3$ TeV
(beyond present LHC top-squark mass bounds).

The benchmark point shown in Table \ref{tab:bm} thus takes as parameter
choices
\be
m_0(i)=5\ {\rm TeV},\ m_{1/2}=1.2\ {\rm TeV},\ A_0=-1.6 m_0,\ \tan\beta =10\
{\rm with}\ \mu =250\ {\rm GeV}\ {\rm and}\ m_A=2\ {\rm TeV}.
\ee
It yields $\Delta_{EW}\sim 26$ with $m_h\simeq 125$ GeV
whilst all sparticle masses are beyond present LHC bounds.
The lightest neutralino $\tchi_1^0$ is higgsino-like
with a thermally-produced (TP) relic abundance of
$\Omega_{\tchi}^{TP}h^2\sim 0.016$.
Since we would also like to be natural in the context of the strong
CP problem, we invoke SUSY axions in the DFSZ model and expect the bulk of dark matter to be axions
along with a smattering of higgsino-like WIMPs\cite{Bae:2013bva,Bae:2013hma,Bae:2014rfa}.

To embed light sleptons within natSUSY, we create a model line with variable
third generation MSSM soft mass $m_{E_3}$. Then, dialing $m_{E_3}$
down in value, we can generate light right-slepton masses as shown in
Table \ref{tab:bm}, where we take $m_{E_3}=1.11$ TeV which then generates
a light tau-slepton with mass $m_{\ttau_1}=378$ GeV.
(We use ISAJET v7.91\cite{Paige:2003mg} to generate the SUSY spectrum.)
The lightest slepton eigenstate is given by\cite{Baer:2006rs}
$\ttau_1=\cos\theta_\tau \ttau_L-\sin\theta_\tau \ttau_R$.
The mixing angle $\theta_\tau=89.9^\circ$ listed in Table \ref{tab:bm}
shows that $\ttau_1$ is dominantly $\ttau_R$.
%
\begin{table}[h!]
\centering
\begin{tabular}{lc}
\hline
parameter & $\ttau_1$BM point \\
\hline
$m_0$      & 5 TeV \\
$m_{1/2}$      & 1.2 TeV \\
$A_0$      & -8 TeV \\
$\tan\beta$    & 10  \\
$m_{E_3}$   & 1.11 TeV \\
\hline
$\mu$          & 250 GeV  \\
$m_A$          & 2 TeV \\
\hline
$m_{\tilde{g}}$   & 2826 GeV \\
$m_{\tilde{u}_L}$ & 5458 GeV \\
$m_{\tilde{u}_R}$ & 5484 GeV \\
$m_{\tilde{e}_R}$ & 4954 GeV \\
$m_{\tilde{t}_1}$& 1517 GeV \\
$m_{\tilde{t}_2}$& 3947 GeV \\
$m_{\tilde{b}_1}$ & 3987 GeV \\
$m_{\tilde{b}_2}$ & 5323 GeV \\
$m_{\tilde{\tau}_1}$ & 378 GeV \\
$m_{\tilde{\tau}_2}$ & 5054 GeV \\
$m_{\tilde{\nu}_{\tau}}$ & 5061 GeV \\
$m_{\tilde{\chi}_1^\pm}$ & 261.4 GeV \\
$m_{\tilde{\chi}_2^\pm}$ & 1019.0 GeV \\
$m_{\tilde{\chi}_1^0}$ & 248.0 GeV \\ 
$m_{\tilde{\chi}_2^0}$ & 259.1 GeV \\ 
$m_{\tilde{\chi}_3^0}$ & 539.3 GeV \\ 
$m_{\tilde{\chi}_4^0}$ & 1034.6 GeV \\ 
$m_h$       & 125.0 GeV \\ 
\hline
$\Omega_{\tilde{\chi}_1}^{std}h^2$ & 0.016 \\
$BR(b\to s\gamma)\times 10^4$ & $3.1$ \\
$BR(B_s\to \mu^+\mu^-)\times 10^9$ & $3.8$ \\
$\sigma^{SI}(\tilde{\chi}_1^0, p)$ (pb) & $2.2\times 10^{-9}$ \\
$\sigma^{SD}(\tilde{\chi}_1^0, p)$ (pb)  & $2.9\times 10^{-5}$ \\
$\langle\sigma v\rangle |_{v\to 0}$  (cm$^3$/sec)  & $1.3\times 10^{-25}$ \\
$\Delta_{\rm EW}$ & 26.4 \\
\hline
$\theta_\tau$ & $89.9^\circ$ \\
\hline
\end{tabular}
\caption{Input parameters (TeV) and masses (GeV)
for the light stau natural SUSY benchmark point from the NUHM2+E3 model
with $m_t=173.2$ GeV using Isajet 7.91~\cite{Paige:2003mg}.
}
\label{tab:bm}
\end{table}

\begin{figure}[htb!]
\centering
    {\includegraphics[height=0.4\textheight]{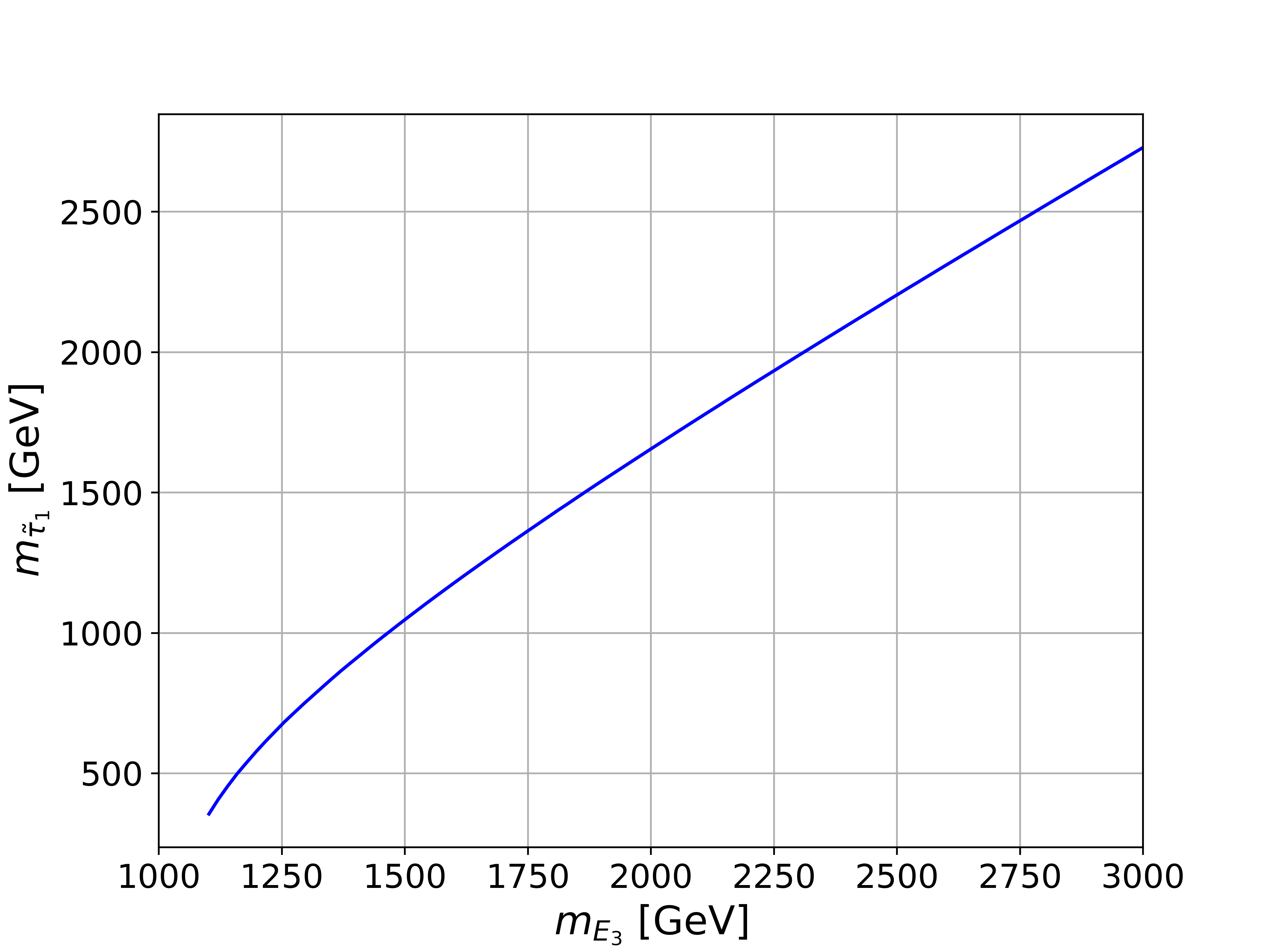}}
    \caption{Plot of $m_{\ttau_1}$ vs. $m_{E_3}$ along the light stau
      natural SUSY model line.
        \label{fig:mline}}
\end{figure}

A plot of $m_{\ttau_1}$ vs. $m_{E_3}$ is shown in Fig. \ref{fig:mline}
for the benchmark point but with variable $m_{E_3}$. The curve cuts off
below $m_{E_3}\alt 1.1$ TeV in that $m_{E_3}^2$ is driven to tachyonic values
at an intermediate iteration in the SUSY RGE solution in ISASUGRA\cite{Baer:1994nc}.
The large $S$ term (defined below Eq. \ref{eq:mE3}) becomes large positive for
non-universal Higgs models with $m_{H_u}\gg m_{H_d}$, which then drives
$m_{E_3}^2$ tachyonic for small enough GUT scale values of $m_{E_3}$.
($S=0$ in models such as CMSSM with universal scalar masses.)
We find this same behavor occurs also in SOFTSUSY\cite{Allanach:2001kg}.
Hence to obtain smaller values of $m_{\ttau_1}$, we implement the weak-scale SUSY
parameters from the BM point into the pMSSM solution embedded in
ISASUSY\cite{Baer:1993ae},
which doesn't include RG running and so allows lighter tau-sleptons
as light as $m_{\ttau_1}\simeq m_{\tchi_1^0}$.

\section{Stau pair production at LHC14}
\label{sec:prod}

\begin{figure}[htb!]
\centering
    {\includegraphics[height=0.4\textheight]{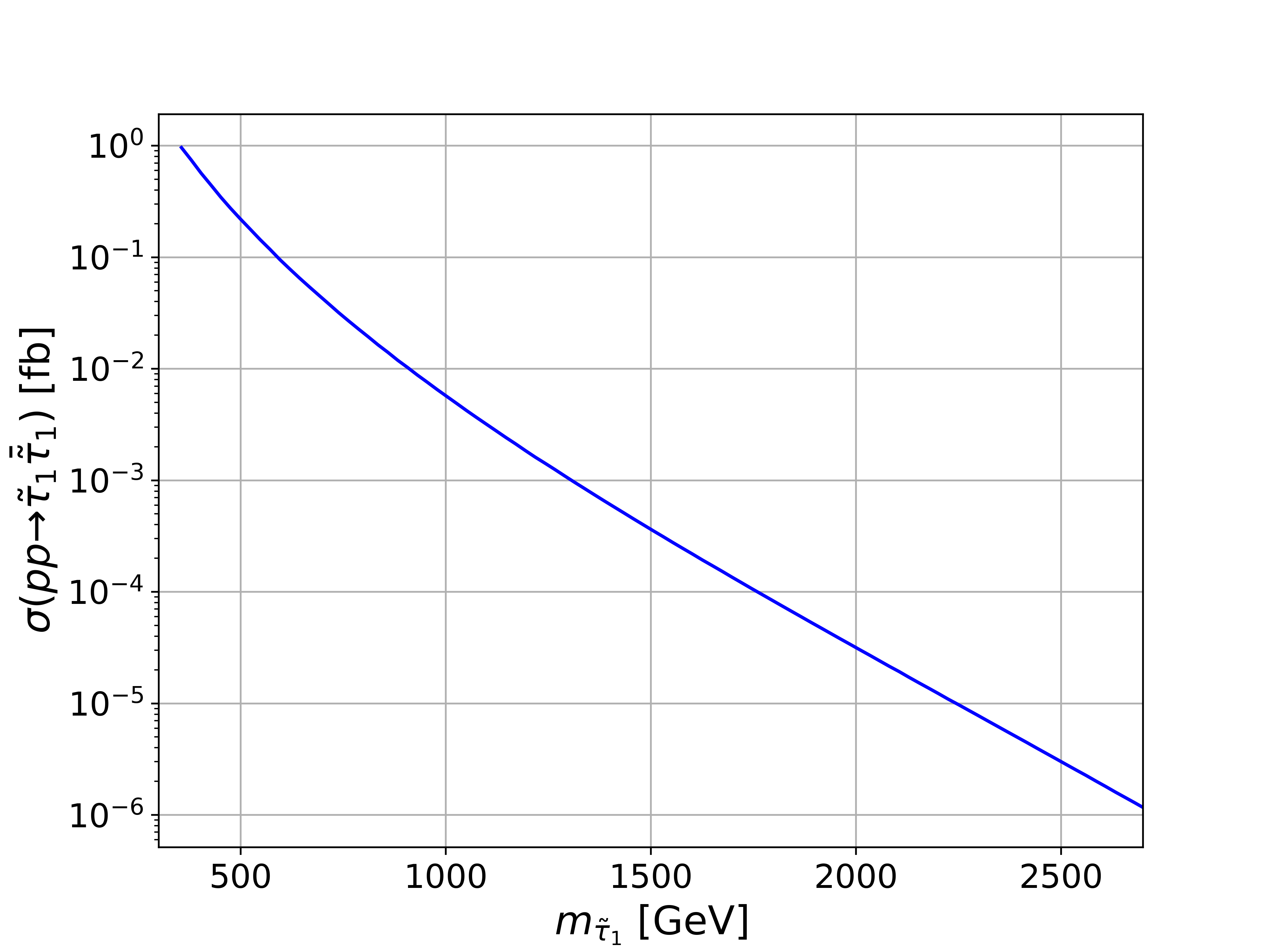}}
    \caption{NLO cross sections (in fb) for $pp\to\ttau_1\ttau_1^* X$
      production at a $pp$ collider with $\sqrt{s}=14$~TeV versus
      $m_{\ttau_1}$ for the
      light $\ttau_1$ natural SUSY model line of the text.
        \label{fig:prod}}
\end{figure}

Pair production of light right-staus takes place via
$q\bar{q}\to\gamma^*,Z^*\to\ttau_1\bar{\ttau}_1X$ at the LHC.
(Light left-staus can also be produced via
$q\bar{q}^\prime \to W^*\to\ttau_1\tnu_{\tau}$.)
Next-to-leading order QCD corrections were computed in Ref. \cite{Baer:1997nh}
and are included in PROSPINO\cite{Beenakker:1996ed} which we use for the
total cross section computation.
The total cross section in fb for production of tau-sleptons at LHC
with $\sqrt{s}=14$ TeV is shown vs. $m_{\ttau_1}$ in Fig. \ref{fig:prod}.
From the plot, we see that $\ttau_1\bar{\ttau}_1$ production occurs
at $\sigma >1$ fb for $m_{\ttau_1}\alt 400$ GeV. For HL-LHC with
an assumed integrated luminosity of 3000 fb$^{-1}$, we would drop below
the 30 total event level for $m_{\ttau_1}\agt 850$ GeV level.
Thus, we would expect any sensitivity of HL-LHC to tau-slepton pair production
to lie in the few hundred GeV region, based solely on total production
cross section.

\section{Right-stau branching fractions in natural SUSY}
\label{sec:decay}

\begin{figure}[htb!]
    \centering
    {\includegraphics[height=0.4\textheight]{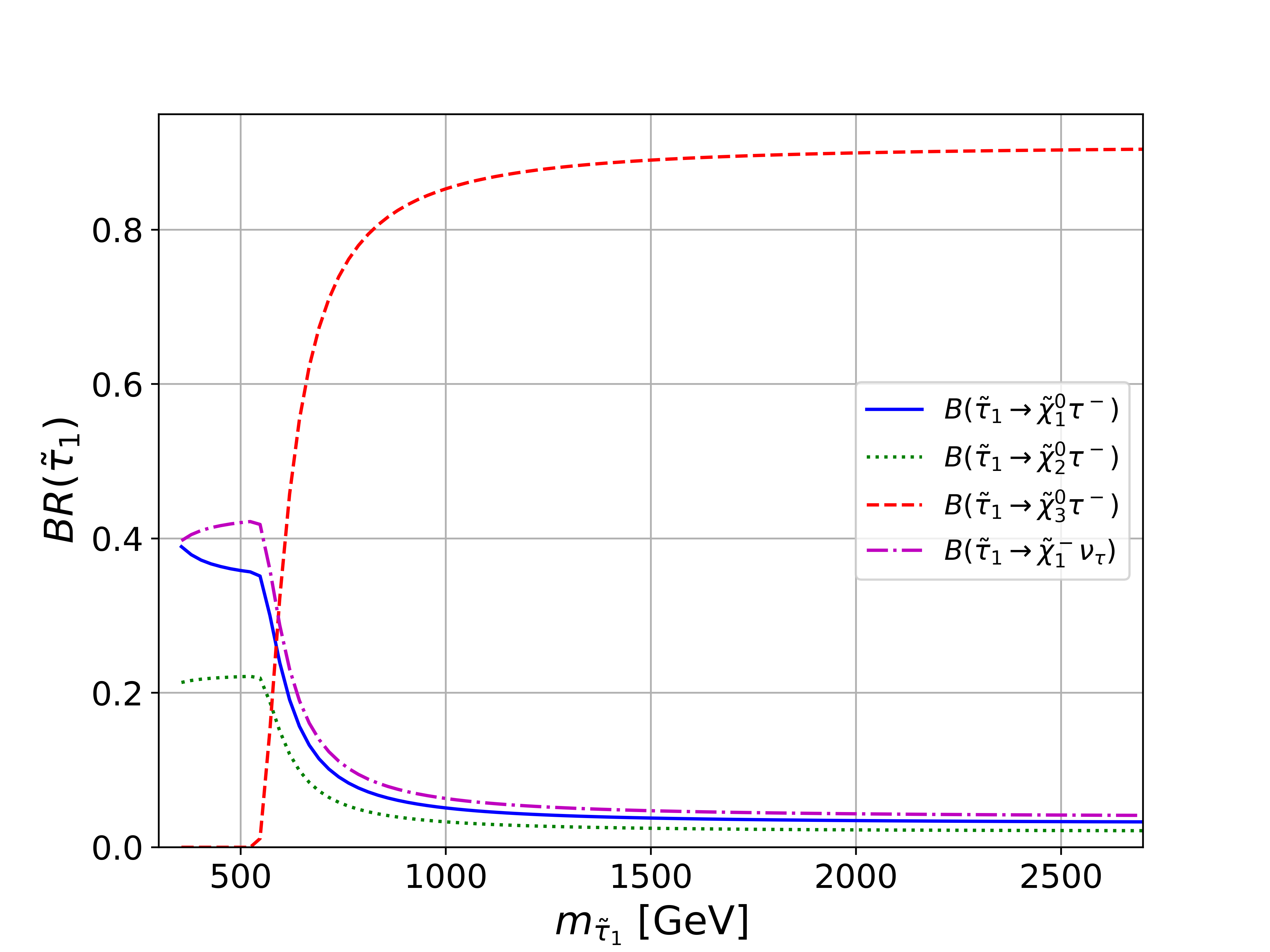}}\\
    \caption{Branching fractions of $\ttau_1$ from natural SUSY
      versus $m_{\ttau_1}$ for the light stau natural SUSY model line.
      \label{fig:BFs}}
\end{figure}

In this Section, we examine the expected light slepton branching
fractions (BFs) within the context of natSUSY. The BFs of
$\ttau_1$ are computed using ISAJET 7.91. In Fig. \ref{fig:BFs},
we plot the dominant $BF(\ttau_1 )$ vs. $m_{\ttau_1}$ along
our $\ttau_1$ natSUSY model line with rather light higgsinos.
From the plot, we see that for $m_{\ttau_1}\alt 550$ GeV, then the decay
$\ttau_1\to\tchi_1^-\nu_\tau$ is actually dominant at $\sim 40\%$.
Since this decay would be followed by $\tchi_1^-\to f\bar{f}^\prime\tchi_1^0$,
with $m_{\tchi_1^+}$ just a few GeV heavier than $m_{\tchi_1^0}$, then
very soft visible energy will ensue and the decay mode is likely to be
hardly visible in the LHC detector environment.

The next largest BF comes from $\ttau_1\to\tchi_1^0\tau$ (blue curve)
which occurs typically at the $\sim 35\%$ level for $m_{\ttau_1}\alt 550$ GeV.
For large enough $\ttau_1 -\tchi_1^0$ mass gap, then this mode can give rise to
visible isolated 1- and 3-prong $\tau$-jets.
The green curve shows the decay $\ttau_1\to\tchi_2^0\tau$,
where the $\tchi_2^0$ is also mainly higgsino-like, but now can decay
as $\tchi_2^0\to f\bar{f}\tchi_1^0$. Again, for small
$\tchi_2^0 - \tchi_1^0$ mass gap, this decay will typically
yield only soft visible energy unless the $\tchi_2^0$ is somewhat boosted.
The $\tau$ lepton may again be visible as a distinctive $\tau$-jet.
Thus, along the model line, and for $m_{\ttau_1}\alt 550$ GeV, we
expect stau pair production to yield either one or two hard $\tau$-jets
plus missing energy, along with possibly soft visible debris from the
quasidegenerate heavier higgsino decays. This is at odds with
simplified model analyses, which usually assume 100\% stau decay
to hard visible $\tau$-jets.

Of further note in Fig. \ref{fig:BFs} is that as $m_{\ttau_1}$ exceeds
the bino mass, where $\tchi_3^0$ is dominantly bino-like,
then new lucrative decay modes open up and rapidly dominate the
$\ttau_1$ branching fractions. For $m_{\ttau_1}\agt 550$ GeV, then
$\ttau_1\to\tchi_3^0\tau$, but the bino $\tchi_3^0 \to\tchi_1^\pm W^\mp$
at about 25\% each, and also $\tchi_3^0\to \tchi_{1,2}^0 Z$ and
$\tchi_1^0 h$ at $\sim 22-25\%$. For this case, then stau pair production
will yield final state events with two hard $\tau$ jets along with
$WW$, $WZ$, $Wh$, $ZZ$, $Zh$ and $hh$.
Such signatures would be very distinctive,
but since the total production cross section tends to be rather low for
such large values of $m_{\ttau_1}$, it is unlikely that these
would be easily visible at HL-LHC with $\sqrt{s}=14$ TeV.

\section{Reach of HL-LHC for stau pairs in natural SUSY}
\label{sec:reach}

In this Section, we examine the potential of HL-LHC ($pp$ collisions at
$\sqrt{s}=14$ TeV with 3000 fb$^{-1}$) for probing $\ttau^+_R\ttau^-_R$
pair production in the context of natural SUSY.

\subsection{Event generation for signal and background}
\label{ssec:evgen}

We use Isajet to first construct a SUSY Les Houches
Accord (SLHA) file\cite{Skands:2003cj} for any natural SUSY
parameter-space point and feed this into Pythia\cite{Sjostrand:2006za} to
generate signal events. We also use Pythia to generate the various $2\to
2$ background (BG) processes.  For $2\to 3$ background processes, we use
Madgraph\cite{Alwall:2011uj} coupled with Pythia.  For our computation
of SM backgrounds to the stau-pair signal, we include parton level
production of $t\bar{t}$, $t\bar{t}V$, $V+jets$ and $VV$ production
(here, $V$ stands for $W^\pm$ or $Z$).
Specifically, we normalize the stau pair production cross sections to their
NLO values obtained from Prospino. For the most important SM backgrounds,
we normalize the cross sections to their values at the NLO
level or better when available.
The NNLO/NNLL $t\bar{t}$ cross section is normalized to
985.7~pb,\footnote{This is taken from
  {\it https://twiki.cern.ch/twiki/bin/view/LHCPhysics/TtbarNNLO} where
  references to the literature for the calculation may also be found.}
the cross sections for $t\bar{t}V$ production are from
Ref.\cite{LHCHiggsCrossSectionWorkingGroup:2016ypw}, $V+j$ cross sections are calculated using the $K$-factor from the ratio of NLO and LO cross
sections from MadGraph with parton jets defined using the anti-$k_T$
algorithm with $p_{Tj}>25$~GeV and $\Delta R=0.4$, and finally $VV$
cross sections are normalized using the results in
Ref.\cite{Campbell:2011bn}.
We use the Delphes code\cite{deFavereau:2013fsa} for detector simulation in our
analysis.

Since our discovery channel contains backgrounds with high transverse momentum
$W$ and $Z$ bosons decaying leptonically or hadronically, we focus on
hard leptons and jets in the central part of the detector.
With this in mind, we require isolated electrons and muons to satisfy,
\begin{itemize}
\item $p_T(e)> 20$ ~GeV, $|\eta_e| < 2.47$, with $P_{TRatio}< 0.1$,  and 

\item $p_T(\mu)> 25$ ~GeV,  $|\eta_\mu| < 2.5$ with $p_{TRatio}<0.2$,
\end{itemize}
where $P_{TRatio}$ is defined as the ratio of the transverse momentum $(p_T^\ell)$ of the lepton to the scalar sum of the transverse momenta of all other particles in a $\Delta R=0.3$ cone around the lepton: $P_{TRatio} \equiv \frac{p_T^\ell}{\sum_{i\in \text{cone}}p_T^i}$.

We construct jets using an anti-$k_T$ jet algorithm and require,
\begin{itemize}
\item  $p_T(j)>20$~GeV with a cone size $R \le 0.4$ and
$|\eta(j)|<4.5$. 
\end{itemize}
A jet is labeled as a $b$-jet if, in addition, it is tagged as a $b$-jet by Delphes. 

For our signal search, we require additional triggers to select candidates
events.
A hadronic $\tau$ jet $\tau_h$ satisfies
\begin{enumerate}
\item the requirement of a baseline jet,
\item $|\eta_j|< 2.4$ and
\item be tagged as a $\tau$-jet by Delphes\footnote{Efficiency and mistag rate
  taken from Ref. \cite{ATLAS:2022aip} (loose working point).
  For 1-prong, the efficiency is set to 85\%.
  For 3-prong, it is 75\%.}
\end{enumerate}

\subsection{$\tau_h\tau_h+\eslt$ signal cuts}
\label{ssec:hh}

\begin{figure}[htb!]
    \centering
    {\includegraphics[height=0.4\textheight]{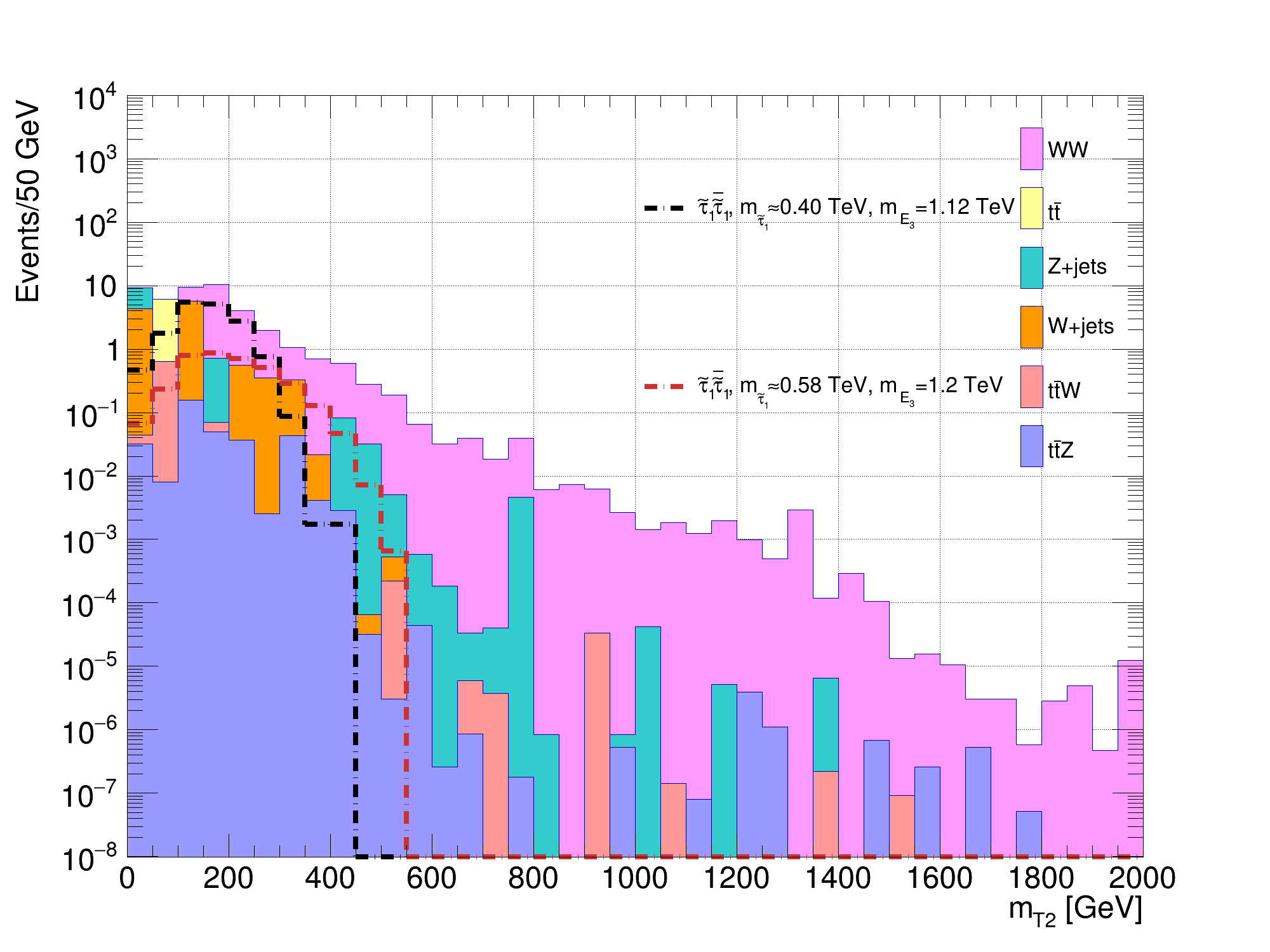}}\\
    \caption{Distributions in $m_{T2}$ for $\tau_h\tau_h+\eslt$
      events from several right-stau pair production models and
      SM backgrounds at HL-LHC where we assume 3000 fb$^{-1}$
      of integrated luminosity.
      \label{fig:mT2_hh}}
\end{figure}

For this (dominant) signal channel, and after scrutinizing various
signal and BG distributions, we require 
\bi
\item At least two OS $\tau_h$ which satisfy the SR $\tau_h$
jet candidate requirement for signal search, $p_T(\tau_1) > 115$ GeV, and $p_T(\tau_2) > 60$ GeV for the two $\tau_h$ selected as candidates.
\ei
Then we require the following cuts:
\bi
\item $n(b) = 0$,
\item $\eslt > 100$ GeV,
\item $\eslt_{,rel}:=$
  $\eslt\cdot\sin{(min(\Delta\phi,\frac{\pi}{2}))}>100$ GeV, where
  $\Delta\phi$ is the azimuthal angle between the $\Vec{\eslt}$ and the
  closest lepton or jet with $p_T > 25$ GeV,
\item $\eslt /\sqrt{H_T} > 5.5$,
\item $|\eta(\tau_1)-\eta (\eslt )| < 4.3$,
\item $m_T(\tau_1,\eslt )+m_T(\tau_2,\eslt ) > 350$ GeV,
\item $min(m_T(\tau_1),\eslt), m_T(\tau_2,\eslt)) > 105$ GeV,
\item $\Delta\phi(\tau_1,\vec{\eslt}) > 55^\circ$,
\item $\Delta\phi (\tau_1, \tau_2) > 50^\circ$ and 
\item $R(\tau_1, \tau_2) < 3.3$.
\ei
After these cuts, we next plot the $m_{T2}$ distribution\cite{Barr:2003rg}
for the $\tau_h\tau_h +\eslt$ events for signal and SM BGs.
The results are shown in Fig.~\ref{fig:mT2_hh}.
The colored solid histograms show various SM backgrounds of which $WW$ is
dominant for $m_{T2}\agt 100$ GeV.
Two signal BM points with various assumed $m_{\ttau_1}$ values are shown
as dot-dashed histograms.
From the plot, we see that the signal histogram for $m_{\ttau_1}=400$
GeV is comparable to the summed background for $m_{T2}\sim 100-300$ GeV
while the $m_{\ttau_1}=580$ GeV signal distribution lies well below BG.
For the $m_{\ttau_1}=400$ GeV BM point, we expect of order tens of
signal events around $m_{T2}\sim 100-200$ GeV
for 3000 fb$^{-1}$ of integrated luminosity. 

\subsection{$\tau_h\ell+\eslt$ signal cuts}
\label{ssec:hl}

For this (subdominant) signal channel, after scrutinizing the
signal and BG distributions, we require the following cuts:
\bi
\item At least one pair of OS lepton and $\tau_h$,
$p_T(\tau_h ) > 165$ GeV
\ei
Then we require
\bi
\item $n(b) = 0$,
\item $\eslt > 100$ GeV,
\item $\eslt_{,rel}:=$
  $\eslt\cdot\sin{(min(\Delta\phi,\frac{\pi}{2}))}>100$ GeV, where
  $\Delta\phi$ is the azimuthal angle between the $\Vec{\eslt}$ and the
  closest lepton or jet with $p_T > 25$ GeV,
\item $|\eta(\ell )| < 2$,
\item $m_T(\tau,\eslt )+m_T(l,\eslt) > 425$ GeV,
\item $\Delta\phi ([\tau_h +\ell], \eslt) > 150^\circ$,
\item $m_T(\tau, \eslt) > 145$ GeV and 
\item $R(\tau_h ,\ell) < 3.1$.
\ei

\begin{figure}[htb!]
    \centering
    {\includegraphics[height=0.4\textheight]{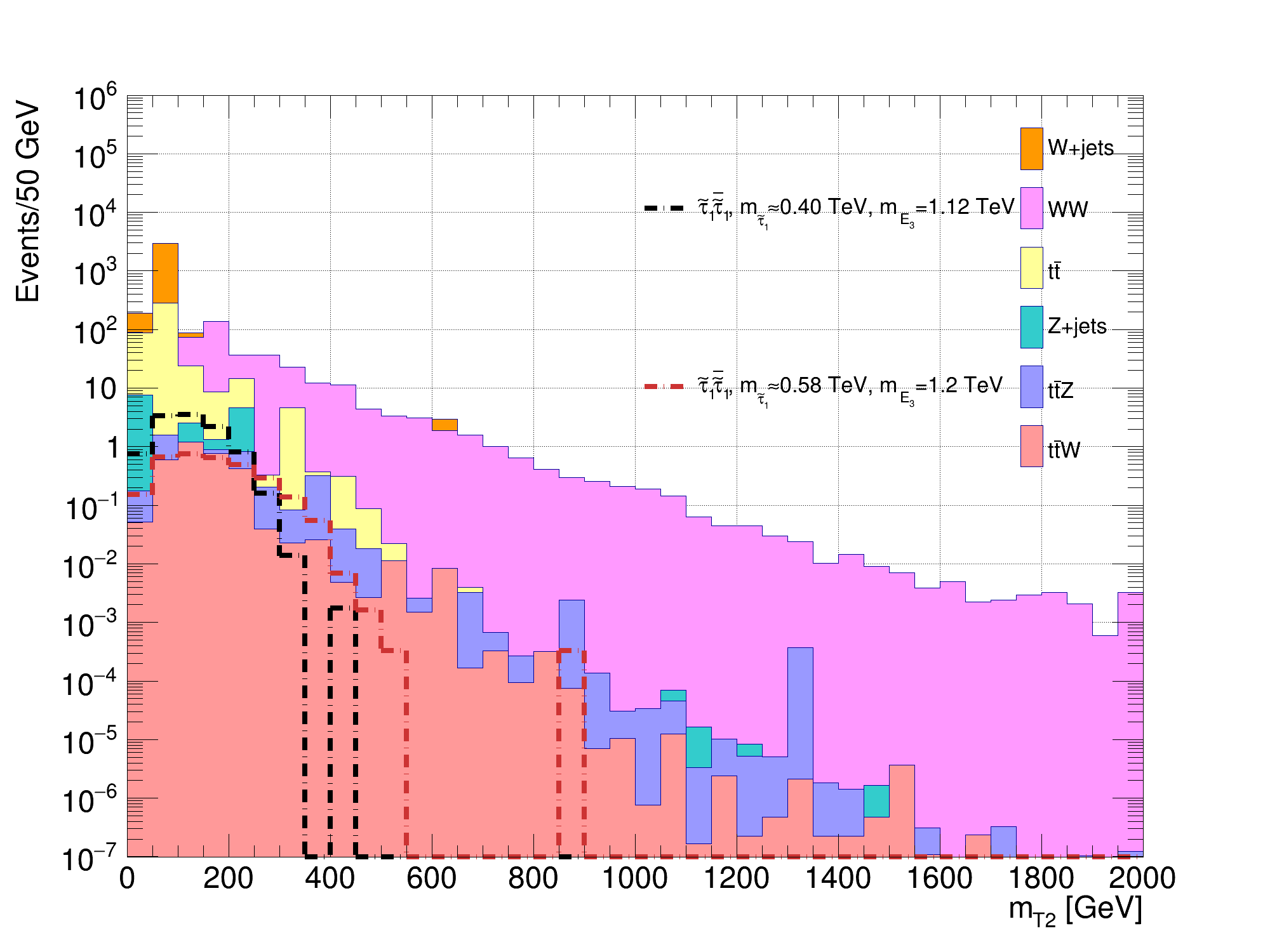}}\\
    \caption{Distributions in $m_{T2}$ for $\tau_h\ell^\pm +\eslt$
      events from several right-stau pair production models and
      SM backgrounds at HL-LHC.
      \label{fig:mT2_hl}}
\end{figure}

The resulting distributions in $m_{T2}$ are shown in Fig. \ref{fig:mT2_hl}
with color-coding as in Fig. \ref{fig:mT2_hh}.
In this case, we find the signal histograms to be well below BG by
at least an order of magnitude even in the most propitious bins.

\subsection{Reach of HL-LHC for stau pair production}
\label{ssec:reach}

For each of the two signal channels from Subsec's \ref{ssec:hh} and \ref{ssec:hl},
we examine the binned $m_{T2}$ distributions
shown in Fig.~\ref{fig:mT2_hh}-\ref{fig:mT2_hl}.
For exclusion of the stau-pair signal, we assume that
the true distribution we would observe in an experiment would correspond
to a background only distribution. Upper limits on $m_{\ttau_1}$ are then
evaluated using a modified frequentist $CL_S$ method \cite{Read_2002}
with the profile likelihood ratio as the test statistic.  The likelihood
is built as a product of Poissonian terms for each of the bins in the
distributions. A background systematic uncertainty is accounted for by
introducing an independent nuisance parameter for each bin of each
channel and the likelihood is modified by log-normal terms to account
for these nuisance parameters, with uncertainty that we take to be 25\%.
Then, the largest value of $m_{\ttau_1}$  that can be excluded at 95\%CL for a given assumed value of
$m_{\tchi_1^0}$ is the exclusion limit.
For discovery, we assume that the distribution one would observe in an
experiment corresponds to signal-plus-background.
We then test this against the background only distribution for
each value of $m_{\ttau_1}$. If the background only hypothesis can be rejected at at least the
5$\sigma$ level, we deem that the HL-LHC would discover staus with a
mass corresponding to that choice of $m_{\ttau_1}$.
For both the exclusion and discovery limits, we use the asymptotic
expansion for obtaining the
median significance \cite{Cowan_2011}.\footnote{We have checked that for
  every channel that we study there are at least ten (frequently
  significantly more) background events in the ``sensitive regions'' of
  the histograms in Fig.~\ref{fig:mT2_hh}-\ref{fig:mT2_hl}. This is
  large enough to justify the use of asymptotic formulae since for
  discovery (exclusion) we are concerned with fluctuations of the
  background (signal plus background).}

\begin{figure}[htb!]
    \centering
    {\includegraphics[height=0.4\textheight]{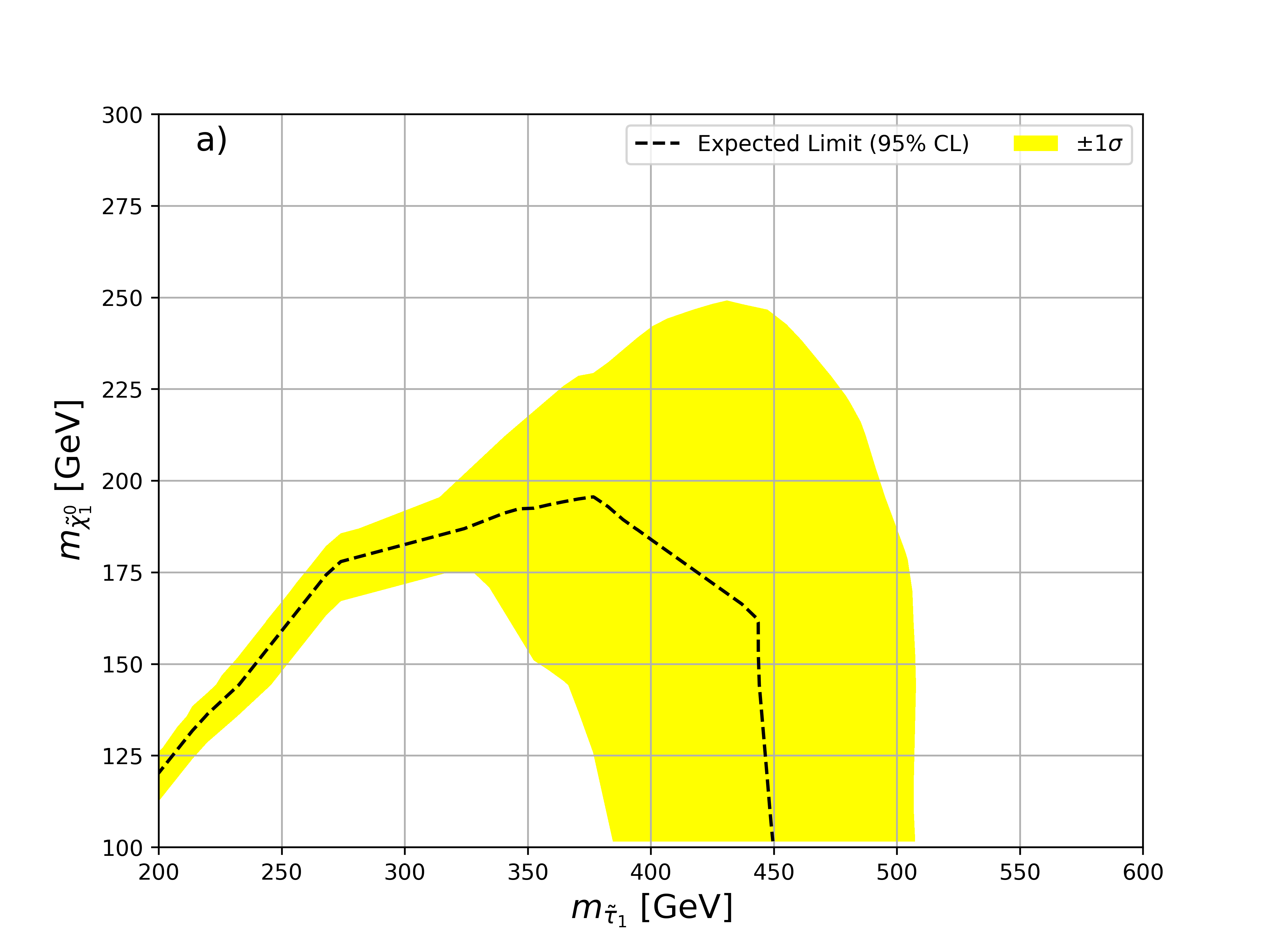}}\\
    {\includegraphics[height=0.4\textheight]{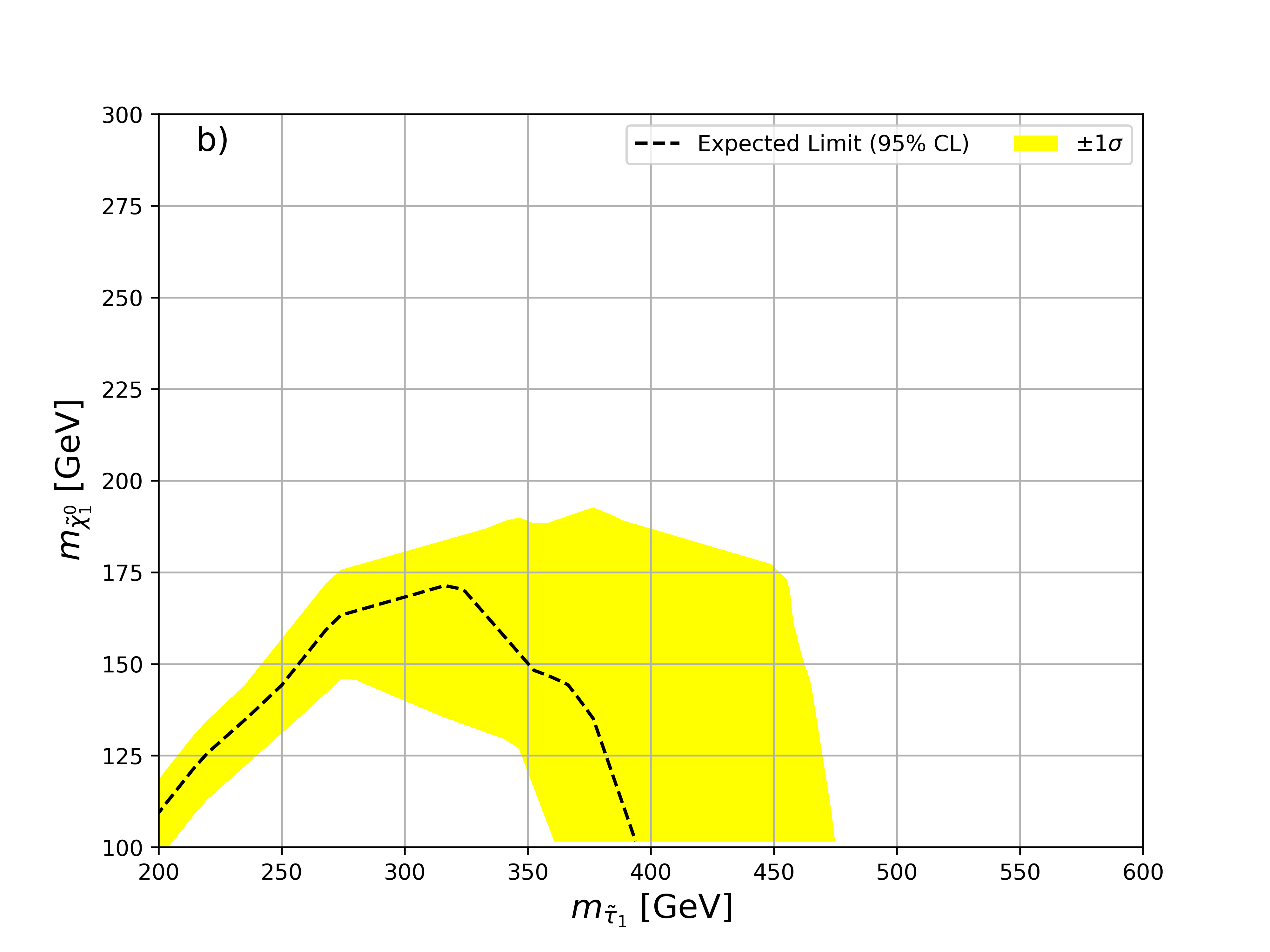}}\\
    \caption{The 95\% CL exclusion reach of HL-LHC for
      right-stau pair production events with a natSUSY setup.
      In frame {\it a}), we assume no systematic error whilst in frame {\it b})
      we assume 25\% systematic error.
      \label{fig:reach}}
\end{figure}

Our HL-LHC reach results are shown in Fig. \ref{fig:reach}{\it a})
in the $m_{\ttau_1}$ vs. $m_{\tchi_1^0}$ plane assuming the
natSUSY benchmark scenario for the cse of no assumed systematic error.
We vary $\mu$ in order to vary $m_{\tchi_1^0}$.
In our case of $\ttau_1\bar{\ttau}_1$ production within natSUSY,
we do not find any discovery reach. However, the 95\% CL exclusion curve
is shown as the black dashed curve along with $1\sigma$ fluctuation limits
shown as the yellow band.\footnote{For Fig. \ref{fig:reach}{\it a}),
  the yellow band is purely statistical uncertainty.
  For Fig. \ref{fig:reach}{\it b}), the band includes the combined
  effects of statistical and systematic uncertainty.
  The reasons leading to larger (statistical) uncertainties for heavier
  mass scales of staus or neutralinos region are
  1) Signal yield is small when $m_{\ttau_1}$ is large and
  2) When the $m_{\tchi_1^0}$ is heavier, the mass difference between the
    stau and the $\tchi_1^0$ becomes smaller such that the visible decay
    products become softer and thus the signal signatures are less
    distinguishable from the backgrounds.
    In both cases, the statistical uncertainty tends to be large which
    widens the uncertainty band. Such features are consistent with the
    reach contour from the current ATLAS search on direct stau production
    via $\tau_h\tau_h+\eslt$.
    See {\it e.g.} Fig. 7 of \cite{ATLAS:2019gti}.}
Unlike the ATLAS and CMS results, our $m_{\tchi_1^0}$ values only extend
down to $\sim 100$ GeV since LEP2
is expected to exclude higgsino-like charginos with mass
$m_{\tchi_1^\pm}\alt 100$ GeV. For $m_{\tchi_1^0}\sim 100$ GeV, then we
expect LHC experiments to be able to exclude $m_{\ttau_1}:200-450$ $(400)$ GeV, assuming $0\%$ $(25\%)$ systematic uncertainty.
For lower values of $m_{\ttau_1}\alt 200$ GeV, then the final state $\tau$s
become too soft for our cuts, while for $m_{\ttau_1}\agt 450$ GeV, then the
expected signal rates become too tiny for exclusion. We see that we
do expect some exclusion for $m_{\tchi_1^0}$ values as high as $\sim 200$ GeV;
for higher $m_{\tchi_1^0}\sim \mu \agt 200$ GeV values, then most of the
final-state energy goes into making to $\tchi_1^0$ rest mass, and too little
visible energy is left to distinguish a signal.
For our exclusion plot, the above cuts were optimized for $\mu\sim 200$ GeV,
so some small extension of this region may be gained if a lighter value of
$\mu$ is assumed (but then one may begin to conflict with
ATLAS/CMS bounds on $\mu$ from soft isolated dilepton plus jets plus
$\eslt$ search results\cite{ATLAS:2019lng,CMS:2021edw}).
In frame Fig. \ref{fig:reach}{\it b}), we show how the reach is diminished
if instead we include an assumed 25\% systematic error.

Comparing our results to ATLAS and CMS, we find in the ATLAS (2018)
Ref. \cite {ATLAS:2018diz} HL-LHC reach study that there also is no
$5\sigma$ discovery reach for $\ttau_R\bar{\ttau}_R$ pairs, although there is a
95\% CL exclusion region for $m_{\tchi_1^0}\alt 100$ GeV.
This study includes some systematic errors plus an assumed pile-up
that we have not included.
This helps us to gain a larger exclusion region than ATLAS even though
some of our staus decay invisibly.

Comparing with the CMS HL-LHC reach study, they do obtain a $5\sigma$
discovery reach even for $m_{\tchi_1^0}>100$ GeV.
There reach should be better than our since they include both
$\ttau_L\ttau_L^*$ and $\ttau_R\ttau_R^*$ production (but see the discussion
below on this dangerous assumption).
Meanwhile, their 95\% CL exclusion limit ranges
from $m_{\ttau_1}\sim 100-650$ GeV for $m_{\tchi_1^0}\sim 100$ GeV which is
broader than our result although we include the invisible stau decay modes
which are generic for natural SUSY, while they assume $\ttau_1\to\tau\tchi_1^0$
at 100\% branching fraction.

\section{Summary and discussion}
\label{sec:conclude}

We have examined right-stau pair production at HL-LHC in the context of a
natural SUSY model line wherein all independent contributions to the
weak scale are comparable to $m_{weak}$ (thus, no weak scale finetuning is
needed). This class of models can be considered as much more plausible
than finetuned models which require accidentally large cancellations
to obtain $m_{W,Z,h}\sim 100$ GeV.
Furthermore, the right-staus are usually expected to be the lightest of
the tau-sleptons. Thus, we embed light right staus within a natSUSY
model line.

In such models, the four higgsino-like EWinos are the lightest of sparticles,
so $\ttau_1\to \tchi_1^-\nu_\tau$ (nearly invisible) at rates comparable
to $\ttau_1\to \tchi_{1,2}^0\tau$. The latter decays lead to ditau $+\eslt$
events at a reduced rate compared to the usual simplified models.
The hadronic ditau$+\eslt$ ends up being a more lucrative search channel
than $\tau_h\ell+\eslt$. By computing signal and SM BG in the $m_{\ttau_1}$ vs.
$m_{\tchi_1^0}$ plane, we do not find any $5\sigma$ discovery regions at
HL-LHC, but we do obtain a 95\%CL exclusion reach. This region extends from
$m_{\ttau_1}:200-450$ $(400)$ GeV, assuming $0\%$ $(25\%)$ systematic uncertainty, for $m_{\tchi_1^0}\sim 100$ GeV, but disappears
entirely for $m_{\tchi_1^0}\agt 200$ GeV.
The net reach is of course reduced by including an overall systematic error.
These results illustrate the difficulty of finding light tau-sleptons at
HL-LHC in a natSUSY context.

\subsection{Comparison of stau pair searches in natural and unnatural SUSY}

Most experimental search projections occur within simplified models which
assume stau pair production ($\ttau_L\ttau_L^*$ and/or $\ttau_R\ttau_R^*$)
along with decay to a single light neutralino $\tchi_1^0$ with
$BF(\ttau_i\to \tau\tchi_1^0)$ at 100\%. Can one distinguish these
presumably unnatural models (with decoupled higgsinos so that $\mu$ must be
large) from our case of natSUSY which includes light higgsinos since
$|\mu|\alt 350$ GeV?
Most likely, the answer is yes.
The natural case with light higgsinos will be accompanied by higgsino
pair production signals\cite{Baer:2011ec}
such as $pp\to \tchi_1^0\tchi_2^0$ and $\tchi_1^\pm\tchi_2^0$ with
$\tchi_2^0\to\ell\bar{\ell}\tchi_1^0$. Higgsino pair production thus gives
rise to soft opposite-sign dileptons $+\eslt$ which may be visible if
the soft leptons recoil against a hard initial state jet
radiation\cite{Han:2014kaa,Baer:2014kya,Baer:2021srt}.
In fact, there are some small excesses in both ATLAS\cite{ATLAS:2019lng}
and CMS\cite{CMS:2021edw} Run 2 data in this channel.
There is also a distinctive same-sign diboson signature in natSUSY
which can occur from wino pair production followed by decay to
lighter higgsinos\cite{Baer:2013yha,Baer:2017gzf}.
These higgsino related signatures would not occur for unnatural SUSY models.
For a review of collider signals from natural SUSY, see {\it e.g.} the
review \cite{Baer:2020kwz}.

\subsection{The case of left- vs. right staus}

The projected reach of ATLAS\cite{ATLAS:2018diz} and CMS\cite{CMS:2018imu} has been computed for stau pair production
followed by 100\% branching fraction $\ttau_i\to \tau\tchi_1^0$ where
$i=L$ and/or $R$. Another question arises: can one tell whether one
is producing $\ttau_L\ttau_L^*$ from $\ttau_R\ttau_R^*$ production, or
indeed a mixture of both as is assumed in some experimental simplified
model scenarios. One way may be to use the different decay energy distrbutions
from left- versus right-tau lepton decays that arise from their parent
$\ttau_L$ or $\ttau_R$ particles.

So far, we have argued that the lighter right-stau
$m_{\ttau_R}\ll m_{\ttau_L}$ is more theoretically motivated, and so we have
focussed on his case. In fact, the ATLAS and CMS studies assume certain
simplified models which violate major theoretical constraints.
Aside from assuming 100\% stau branching fractions into a single
mode $\tau\tchi_1^0$, if one assumes a light left-stau $\ttau_L$,
then necessarily it comes with a light tau-sneutrino.
The weak scale mass relations are\cite{Baer:2024kms} (neglecting small mixing effects)
\bea
m_{\ttau_L}^2&\simeq &m_{L_3}^2+m_\tau^2+m_Z^2\cos 2\beta (-1/2+\sin\theta_W)\\
m_{\tnu_{\tau L}}^2&\simeq &m_{L_3}^2+m_Z^2\cos 2\beta (+1/2)\ and \\
m_{\ttau_R}^2&\simeq &m_{E_3}^2+m_\tau^2+m_Z^2\cos 2\beta (-\sin\theta_W)
\eea
so that most of the mass of $\ttau_L$ and $\tnu_{\tau L}$ comes from
$m_{L_3}^2$. This actually means that if you assume light left staus,
one must also include $pp\to W^*\to \ttau_L\tnu_{\tau L}$ and $\tnu_{\tau L}\tnu_{\tau L}^*$ production where now the $\tnu_{\tau L}$ also usually decays visibly.
(Even if one assumes an invisible $\tnu_{\tau L}\to\nu_\tau\tchi_1^0$ decay,
the $W^*$ production channel will give rise to a large rate for
mono-tau-jet$+\eslt$ events which should be included in any search strategy.)
The $W^*$ mediated production cross section dominates slepton
pair production\cite{Baer:1993ew} and sneutrino pair production is comparable to
stau-left pair production cross section. Thus, the total cross sections for left-slepton pair production will be much higher than typically assumed in
simplified models, and the decay sigatures will be more complex, for a given value of $m_{\ttau_L}$. We expect this much more complex, but more realistic,
case of left-slepton pair production to be readily distinguishable from
right stau pair production. At present, realistic analyses are lacking.

{\it Acknowledgements:} 

We thank Xerxes Tata for comments on the manuscript.
This material is based upon work supported by the U.S. Department of
Energy, Office of Science, Office of High Energy Physics under Award
Number DE-SC-0009956. VB gratefully acknowledges
support from the William F. Vilas estate.


\bibliography{stau2}
\bibliographystyle{elsarticle-num}

\end{document}